\documentclass[prd, aps, superscriptaddress, preprintnumbers, floatfix, nofootinbib]{revtex4}

\usepackage{longtable}
\usepackage{amsfonts}
\usepackage{amsmath}
\usepackage{amssymb}
\usepackage{bm}
\usepackage{dcolumn}
\usepackage{epsfig}
\usepackage{graphicx}   
\usepackage{graphics}
\usepackage[utf8]{inputenc}
\usepackage{latexsym}
\usepackage{rotating}
\usepackage{hyperref}

\usepackage{amsfonts}
\usepackage{amsmath}
\usepackage{amssymb}
\usepackage{bm}
\usepackage{dcolumn}
\usepackage{epsfig}
\usepackage{graphicx}
\usepackage{graphics}
\usepackage[utf8]{inputenc}
\usepackage{latexsym}
\usepackage{rotating}
\usepackage{hyperref}
\usepackage{color}
\usepackage{soul}
\usepackage[normalem]{ulem}
\newcommand\be{\begin{equation}}
\newcommand\ba{\begin{eqnarray}}
\newcommand\ee{\end{equation}}
\newcommand\ea{\end{eqnarray}}

\begin{document}

\title{Dynamical analysis on $f(R,G)$ cosmology}

\author{S. Santos da Costa}\email{simonycosta@on.br}
\affiliation{Departamento de Astronomia, Observat\'orio Nacional, 20921-400, Rio de Janeiro, RJ, Brasil}
\affiliation{Dipartimento di Fisica ”E. Pancini”, Universit`a di Napoli ”Federico II”, Compl. Univ. di Monte S. Angelo, Edificio G, Via Cinthia, I-80126, Napoli, Italy}

\author{F. V. Roig}\email{froig@on.br}
\affiliation{Departamento de Astronomia, Observat\'orio Nacional, 20921-400, Rio de Janeiro, RJ, Brasil}

\author{J. S. Alcaniz}\email{alcaniz@on.br}
\affiliation{Departamento de Astronomia, Observat\'orio Nacional, 20921-400, Rio de Janeiro, RJ, Brasil}
\affiliation{Physics Department, McGill University, Montreal, QC, H3A 2T8, Canada}

\author{S. Capozziello}\email{capozziello@na.infn.it}
\affiliation{Dipartimento di Fisica ”E. Pancini”, Universit`a di Napoli ”Federico II”, Compl. Univ. di Monte S. Angelo, Edificio G, Via Cinthia, I-80126, Napoli, Italy}
\affiliation{Istituto Nazionale di Fisica Nucleare (INFN) Sez.  di Napoli, Compl. Univ. di Monte S. Angelo, Edificio G, Via Cinthia, I-80126, Napoli, Italy}
\affiliation{Gran Sasso Science Institute (INFN), Viale F. Crispi, 7, I-67100, L’Aquila, Italy}

\author{M. De Laurentis}\email{laurentis@th.physik.uni-frankfurt.de}
\affiliation{Institute for Theoretical Physics, Goethe University, Max-von-Laue-Str. 1, D-60438 Frankfurt, Germany}
\affiliation{Tomsk State Pedagogical University, 634061 Tomsk, Russia}

\author{M. Benetti} {\email{micolbenetti@on.br}
\affiliation{Departamento de Astronomia, Observat\'orio Nacional, 20921-400, Rio de Janeiro, RJ, Brasil}

\date{\today}

\begin{abstract}
We use a dynamical system approach to study the cosmological viability of $f(R,\mathcal{G})$ gravity theories. The method consists of formulating the evolution equations as an autonomous system of ODEs, using suitable variables. The formalism is applied to a class of models in which $f(R,\mathcal{G})\propto R^{n}\mathcal{G}^{1-n}$ and its solutions and corresponding stability are analysed in detail. New accelerating solutions that can be attractors in the phase space are found. We also find that this class of models does not exhibit a matter-dominated epoch,  a solution which is  inconsistent with current cosmological observations. 

\end{abstract}


\maketitle

\section{Introduction}

In the past decades, our ability to perform high-precision measurements of distance to type Ia supernovae (SNe Ia)~ \cite{SNE, SNE2}, CMB anisotropies~\cite{CMB,CMB2}, the large-scale clustering patterns of galaxies~ \cite{BAO,BAO2,BAO3,BAO4} and the age measurements of high-$z$ galaxies~\cite{age1,age2,age3} led  to the remarkable conclusion that the Universe is currently accelerated. In Einstein's general relativity (GR), such behaviour implies either the existence of a new field, the so-called dark energy, or that the matter content of the universe is subject to dissipative processes (see, e.g., \cite{dissipative1, dissipative2, dissipative3}). Among the proposals to address this problem, the one that best fits almost all available observations, i.e., the standard cosmological constant $\Lambda$, is affected by significant fine-tuning problems related to the vacuum energy scale~\cite{weinberg}. Thus, it is of fundamental importance to investigate the cosmological viability of other theoretical scenarios that may explain the current accelerating expansion phase of the universe. 

In one of these proposals, the mechanism behind the cosmic acceleration is attributed to a modification of the standard theory of gravity on cosmologically relevant physical scales, without invoking  the existence of a dark energy field. The simplest family of these extended gravity theories is the so-called $f(R)$ gravity (see, e.g.,~\cite{defelice} and references therein), in which the Einstein-Hilbert action is replaced by a non linear function of the Ricci scalar $R$. Another motivation to study the phenomenon of cosmic acceleration (both in the early and late-time universe) using extended gravity is the possibility of  incorporating quantum corrections to GR in the form of terms of higher order in the curvature. Several combinations of curvature invariants, like $R_{\mu\nu}R^{\mu\nu}$ and $R_{\mu\nu\sigma\rho}R^{\mu\nu\sigma\rho}$, have be considered~\cite{gorbunov, myrzakulov, bamba, sebastiani}, although some of them, e.g. proportional to the Kretschmann invariant ($R_{\mu\nu\sigma\rho}R^{\mu\nu\sigma\rho}$), lead to solutions with increasing spatial anisotropy~\cite{barrow, topo1}, which are incompatible with observations. On the other hand, the Gauss-Bonnet topological invariant $\mathcal{G}$ naturally arises in the process of quantum field theory regularization and renormalization in curved  spacetime~\cite{birrell}. Thus, one can consider a theory where both $R$ and $\mathcal{G}$ ($f(R, \mathcal{G})$) are non linearly present in such a way that we exhaust the budget of curvature degrees of freedom needed to extend GR, since the Ricci scalar and  both the Ricci and the Riemann tensor are present in the definition of $\mathcal{G}$. 

Nevertheless, one of the the main problems in the study of higher order theories of gravity is the extreme difficulty to find exact cosmological solutions, due to the high degree of non linearity exhibited by these theories. This fact makes difficult to obtain both analytical and numerical solutions that can be compared to observations. Consequently, it is important to use different  methods which are able to assist in resolving these problems. One possibility is to use the dynamical system approach (DSA). This approach has the advantage of providing a relatively simple method to obtain numerical solutions, and most importantly, to obtain a qualitative description of the global dynamics of these models. A recent series of papers used the DSA to study the cosmological effects of both $f(R)$ and $f(\mathcal{G})$ models separately (e.g., \cite{Amendola2007, carloni2005, carloni2009} for $f(R)$ and \cite{nojiri, cognola, Li, defelice2009, zhou} for $f(\mathcal{G})$). On the other hand, studies of $f(R,\mathcal{G})$ gravity models using DSA are less common~\cite{ivanov,alimoh}.

We should emphasize, however, that in the literature some works have studied exact cosmological solutions for Gauss-Bonnet gravity which can reproduces the $\Lambda$CDM scenario and/or quintessence behaviour, considering for example explicit choices of the functions $f(\mathcal{G})$, $f(R,\mathcal{G})$ or a scalar field coupled with gravity~\cite{NOS, elizalde_etal}. Further, the stability conditions for different cosmological evolutions in Friedmann-Lemaître-Robertson-Walker (FLRW) universes, such as inflationary epoch and late-time accelerated era as described by the $\Lambda$CDM model, have been studied using a different approach than the DSA~\cite{cruz-dombriz}. Lastly, so important as the stability conditions, it has been verified that Gauss-Bonnet gravity may carry a ghost mode, i.e. a field whose kinetic term in the action is unbounded from below~\cite{calcagni}, in an empty anisotropic universe, the so called Kasner-type background~\cite{felice_tanaka}. However, this problem can be softened in the FLRW limit, mainly if an effective theory describing only small deviation from the FLRW universe and/or the modifications of gravity tend to vanish at early times~\cite{felice_tanaka}.

In this paper, we will apply the DSA to a more general class of theories $f(R,\mathcal{G})$ where, in principle, both  $R$ and $G$ are non linear in the action. These kind of theories have been particularly well studied, mainly motivated by string theory~\cite{metsaev, gasperini1, gasperini2, nojiri2006} and by the Gauss-Bonnet topological invariant, which may solve some shortcomings of the original $f(R)$ gravity and contributes to the accelerated expansion~\cite{defelice2009b, capozziello2014, delaurentis2014, delaurentis2015, eliz, myrzakulov2}. By studying the stability of the solutions and their cosmological viability, we find new accelerating solutions that can be attractors in the phase space. The paper is organized as follows: In section \ref{sec2}, we present the basic equations of the $f(R,\mathcal{G})$ gravity, namely, the action
and the field equations. In section \ref{sec3}, we set up the dynamical system defining suitable variables and deriving their
evolution equations. Then, we analyze the fixed points and their stability for a particular model $f(R,\mathcal{G})=\alpha R^{n}\mathcal{G}^{1-n}$ in section \ref{sec4}. Finally, a discussion on the cosmological viability of this class of theories as well as the main conclusions of our analysis are presented in section \ref{sec5}. 

\section{Basic equations}\label{sec2}

The most general action for the modified Gauss-Bonnet gravity is
\begin{eqnarray}
\mathcal{S}=\frac{1}{2\kappa^2}\int{dx^4\sqrt{-g}\left[f(R,\mathcal{G})+\mathcal{L}_m\right]},
\label{act1}
\end{eqnarray}
where $f(R,\mathcal{G})$ is a function of the Ricci scalar and the Gauss-Bonnet invariant is defined as
\begin{eqnarray}
\mathcal{G}\equiv R^2-4R_{\alpha\beta}R^{\alpha\beta}+R_{\alpha\beta\rho\sigma}R^{\alpha\beta\rho\sigma},
\label{GBterm}
\end{eqnarray}
with $\kappa^2=8\pi G$, $G$ is the Newton constant and $\mathcal{L}_{\mathrm{m}}$ is the standard matter Lagrangian density (we use physical units such that $c=k_{B}=\hbar=1)$.

The variation of the action (\ref{act1}) with respect to the metric field $g_{\mu\nu}$ produces the following field equations
\begin{eqnarray}
R_{\mu\nu}-\frac{1}{2}g_{\mu\nu}R=\kappa^2 T_{\mu\nu}^{m}+T_{\mu\nu}^{GB},\label{ECGB}
\end{eqnarray}
where the stress-energy tensor is defined as usual
\begin{eqnarray}
T_{\mu\nu}^{m}=-\frac{2}{\sqrt{-g}}\frac{\delta (\sqrt{-g}\mathcal{L}_m)}{\delta g^{\mu\nu}},
\end{eqnarray}
and the Gauss-Bonnet tensor is
\begin{eqnarray}
T_{\mu\nu}^{GB}&=&\nabla_{\mu}\nabla_{\nu}f_{R}-g_{\mu\nu}\square f_{R}+2R\nabla_{\mu}\nabla_{\nu}f_{\mathcal{G}}-2g_{\mu\nu}R\square f_{\mathcal{G}}-4R_{\mu}^{\lambda}\nabla_{\lambda}\nabla_{\nu}f_{\mathcal{G}}-
4R_{\nu}^{\lambda}\nabla_{\lambda}\nabla_{\mu}f_{\mathcal{G}}\nonumber \\
&+&4R_{\mu\nu}\square f_{\mathcal{G}}+4g_{\mu\nu}R^{\alpha\beta}\nabla_{\alpha}\nabla_{\beta}f_{\mathcal{G}}
+4R_{\mu\nu\alpha\beta}\nabla^{\alpha}\nabla^{\beta}f_{\mathcal{G}}-\frac{1}{2}g_{\mu\nu}(Rf_{R}+\mathcal{G}f_{\mathcal{G}}-f)\nonumber \\
&+&(1-f_{R})\left(R_{\mu\nu}-\frac{1}{2}g_{\mu\nu}R\right).
\end{eqnarray}
Note that $\square$ is the d'Alembert operator in curved spacetime and here, and henceforth, we use the notations
\begin{eqnarray}
f_{R}\equiv \frac{\partial f(R,\mathcal{G})}{\partial R} \qquad \mbox{and} \qquad f_{\mathcal{G}}\equiv \frac{\partial f(R,\mathcal{G})}{\partial \mathcal{G}},
\end{eqnarray}
for the partial derivatives with respect to $R$ and $\mathcal{G}$.

We consider the FLRW metric for a spatially flat universe ($k=0$) with a time dependent scale factor $a(t)$
\begin{eqnarray}
ds^2=-dt^2+a^2(t)dx^idx_i.
\end{eqnarray}
From this metric, the Ricci scalar and the Gauss-Bonnet invariant are given by
\begin{eqnarray}
R&=&6(2H^2+\dot{H}),\label{RicciScalar}\\
\mathcal{G}&=&24H^2(H^2+\dot{H}),\label{GaussBonnet}
\end{eqnarray}
where $H=\dot{a}/a$ is the Hubble parameter and the dots stands for derivatives with respect to time. Furthermore, the corresponding field equations obtained from \eqref{ECGB} by using the FLRW metric are given by:
\begin{eqnarray}
3f_{R}H^2&=&\kappa^2(\rho_{\mathrm{m}}+\rho_{\mathrm{r}})+\frac{1}{2}(f_{R}R-f-6H\dot{f_{R}}+\mathcal{G}f_{\mathcal{G}}-24H^3\dot{f_{\mathcal{G}}})\label{EC1}
\end{eqnarray}
and 
\begin{eqnarray}
2f_{R}\dot{H}&=&-\kappa^2\left(\rho_{\mathrm{m}}+\frac{4}{3}\rho_{\mathrm{r}}\right)+H\dot{f_{R}}-\ddot{f_{R}}+4H^3\dot{f_{\mathcal{G}}}
-8H\dot{H}\dot{f_{\mathcal{G}}}-4H^2\ddot{f_{\mathcal{G}}}\;,\label{EC2}
\end{eqnarray}
with the matter and radiation densities, $\rho_{\mathrm{m}}$ and $\rho_{\mathrm{r}}$, satisfying the usual continuity equations
\begin{eqnarray}
\dot{\rho}_{\mathrm{m}}+3H\rho_{\mathrm{m}}=0,\\
\dot{\rho}_{\mathrm{r}}+4H\rho_{\mathrm{r}}=0.
\end{eqnarray}

By using a redefinition of the quantities, we can rewrite Eqs. (\ref{EC1}) and (\ref{EC2}) as
\begin{eqnarray}
3f_{R}H^2&=&\kappa^2(\rho_m+\rho_{rad}+\rho_{GDE}),\\
2f_{R}\dot{H}&=&-\kappa^2\left(\rho_m+\frac{4}{3}\rho_{rad}+\rho_{GDE}+p_{GDE}\right),
\end{eqnarray}
so that we have the following identities
\begin{eqnarray}
\kappa^2\rho_{GDE}&=&\frac{1}{2}(f_{R}R-f-6H\dot{f_{R}}+\mathcal{G}f_{\mathcal{G}}-24H^3\dot{f_{\mathcal{G}}}),\\
-\kappa^2(\rho_{GDE}+p_{GDE})&=&H\dot{f_{R}}-\ddot{f_{R}}+4H^3\dot{f_{\mathcal{G}}}
-8H\dot{H}\dot{f_{\mathcal{G}}}-4H^2\ddot{f_{\mathcal{G}}}.
\end{eqnarray}

The energy density $\rho_{\mathrm{GDE}}$ and the pressure density $p_{\mathrm{GDE}}$ of a {\it{geometrical dark energy}} component  defined in this way satisfy the usual conservation equation
\begin{eqnarray}
\dot{\rho}_{GDE}=-3H(\rho_{GDE}+p_{GDE}).
\end{eqnarray}

Hence the equation of state parameter $\omega_{GDE}\equiv p_{GDE}/\rho_{GDE}$ is given by
\begin{eqnarray}
\omega_{GDE}=-1-\frac{H\dot{f_{R}}-\ddot{f_{R}}+4H^3\dot{f_{\mathcal{G}}}
-8H\dot{H}\dot{f_{\mathcal{G}}}-4H^2\ddot{f_{\mathcal{G}}}}{\frac{1}{2}(f_{R}R-f-6H\dot{f_{R}}+\mathcal{G}f_{\mathcal{G}}-24H^3\dot{f_{\mathcal{G}}})},
\end{eqnarray}
and we also define the effective equation of state as
\begin{eqnarray}
\omega_{eff}=-1-\frac{2\dot{H}}{3H^2}.
\end{eqnarray}
In the next section, we will use the equations \eqref{EC1} and \eqref{EC2} to set up the dynamical system analysis.

\section{Dynamical System Approach to $f(R,G)$ Gravity}\label{sec3}

In this section we will use the dynamical system approach developed in \cite{Amendola2007, carloni2005, ivanov, carloni2015} to the cosmology of fourth order gravity. In order to discuss cosmological dynamics for a $f(R,\mathcal{G})$ theory of gravity, we will follow two steps: (i) to introduce convenient dimensionless variables, and (ii) to derive the evolution equations of these variables. The fixed points of this system of equations will thus represent some asymptotic regimes of the Universe evolution.

\subsubsection*{General Formalism}

Let us define the variables:
\begin{eqnarray}
x_1\equiv \frac{\dot{f_R}}{f_RH},\quad x_2\equiv \frac{f}{6f_RH^2},\quad x_3\equiv \frac{R}{6H^2},\quad x_4\equiv \frac{\kappa^2\rho_r}{3f_RH^2},\quad x_5\equiv \frac{\mathcal{G}f_{\mathcal{G}}}{6f_RH^2},\quad x_6\equiv \frac{4H\dot{f_\mathcal{G}}}{f_R},\quad x_7\equiv \frac{\kappa^2\rho_m}{3f_RH^2}. \label{var}
\end{eqnarray}
Thus, from Eq.(\ref{EC1}) we have the algebraic identity
\begin{eqnarray}
1=-x_1-x_2+x_3+x_4+x_5-x_6+x_7,\label{constraint}
\end{eqnarray}
together with the density parameters:
\begin{eqnarray}
\Omega_m\equiv x_7, \qquad \Omega_r\equiv x_4 \qquad \mbox{and} \qquad \Omega_{GDE}\equiv -x_1-x_2+x_3+x_5-x_6
\end{eqnarray}

We also introduce a dimensionless time variable: the logarithmic time $N=\mid\ln{a(t)}\mid$. So, taking the derivative
of these variables with respect to $N$, we obtain the following dynamical system:
\begin{eqnarray}
\frac{dx_1}{dN}&=&\frac{\ddot{f_R}}{f_RH^2}-x_1^2-x_1\frac{\dot{H}}{H^2}\nonumber\\
\frac{dx_2}{dN}&=&\frac{\dot{f}}{6f_RH^3}-x_1x_2-2x_2\frac{\dot{H}}{H^2},\nonumber\\
\frac{dx_3}{dN}&=&\frac{\dot{R}}{6H^3}-2x_3\frac{\dot{H}}{H^2},\nonumber\\
\frac{dx_4}{dN}&=&-2x_3x_4-x_1x_4,\label{sys1}\\
\frac{dx_5}{dN}&=&\frac{\dot{\mathcal{G}}}{\mathcal{G}H}x_5+\frac{\mathcal{G}}{24H^4}x_6-x_1x_5-2x_5(x_3-2),\nonumber\\
\frac{dx_6}{dN}&=&x_6\frac{\dot{H}}{H^2}+4\frac{\ddot{f}_{\mathcal{G}}}{f_R}-x_1x_6,\nonumber\\
\frac{dx_7}{dN}&=&-3x_7-x_1x_7-2x_7\frac{\dot{H}}{H^2}.\nonumber
\end{eqnarray}
In order to close the system, we must have all terms in the right-hand side of the above equations expressed
in terms of variables specified in Eqs. \eqref{var}. So, with the help of the  Eqs. \eqref{RicciScalar}, \eqref{GaussBonnet} and \eqref{EC2}, we find
\begin{eqnarray}
\frac{\dot{H}}{H^2}&=&x_3-2,\\
\frac{\dot{f}}{6f_RH^3}&=&-\frac{x_1x_3}{b},\\
\frac{\dot{R}}{6H^3}&=&\frac{x_1x_3}{b},\\
\frac{\mathcal{G}}{24H^4}&=&x_3-1,\\
\frac{\dot{\mathcal{G}}}{\mathcal{G}H}&=&\frac{1}{(x_3-1)}\left[\frac{x_1x_3}{b}+2(x_3-2)^2\right],\\
\frac{4\ddot{f_{\mathcal{G}}}}{f_R}&=&-3x_7-4x_4+x_1+2\frac{x_1x_3}{b}+x_6(5-2x_3)-2(x_3-2)+\frac{x_5}{x_3-1}\left[2(x_3-2)^2+\frac{x_1x_3}{b}\right]-\frac{\ddot{f_R}}{f_RH^2},
\end{eqnarray}
and the system becomes:
\begin{eqnarray}
\frac{dx_1}{dN}&=&\Gamma -x_1^2-x_1(x_3-2)\nonumber\\
\frac{dx_2}{dN}&=&-\frac{x_1x_3}{b}-x_2(2x_3-4+x_1),\nonumber\\
\frac{dx_3}{dN}&=&\frac{x_1x_3}{b}-2x_3(x_3-2),\nonumber\\
\frac{dx_4}{dN}&=&-2x_3x_4-x_1x_4,\label{sys2}\\
\frac{dx_5}{dN}&=&\frac{x_5}{(x_3-1)}\left[\frac{x_1x_3}{b}+2(x_3-2)^2\right]+x_6(x_3-1)-x_1x_5-2x_5(x_3-2),\nonumber\\
\frac{dx_6}{dN}&=&-x_6(x_1+x_3-3)-3x_7-4x_4-2(x_3-2)+x_1+2\frac{x_1x_3}{b}+\frac{x_5}{x_3-1}\left[2(x_3-2)^2+\frac{x_1x_3}{b}\right]-\Gamma,\nonumber\\
\frac{dx_7}{dN}&=&-x_7(2x_3+x_1-1),\nonumber
\end{eqnarray}
where
\begin{eqnarray}
b\equiv \frac{d\ln{f_R}}{d\ln{R}}=\frac{Rf_{RR}}{f_R} \qquad \mbox{and} \qquad r\equiv -\frac{d\ln{f}}{d\ln{R}}=-\frac{Rf_R}{f}.\label{mer}
\end{eqnarray}
These set of equations describe the cosmological evolution of a general $f(R,\mathcal{G})$ theory of gravity, 
where $\Gamma\equiv\frac{\ddot{f_R}}{f_RH^2}$ specifies the theory. We also define the following expressions:
\begin{eqnarray}
\omega_{eff}&=&-\frac{1}{3}(2x_3-1), \\
\omega_{GDE}&=&-1-\frac{1}{3}\left\{\frac{3x_{7}+4x_{4}+2(x_{3}-2)-2\frac{x_{1}x_{3}}{b}-\frac{x_{5}}{(x_{3}-1)}\left[2(x_{3}-2)^2+\frac{x_{1}x_{3}}{b}\right]}{1-x_{4}-x_{7}}\right\}.
\end{eqnarray}
\vspace{0.5cm}

In general, the system is not closed unless $\Gamma$ is expressed in terms of the dynamical variables \eqref{var}. 
In the next section, we will consider a particular case for $f(R,\mathcal{G})$ and will study its dynamics and stability in a flat FLRW universe.

\vspace{0.5cm}

\section{Power Law $f(R,\mathcal{G})$}\label{sec4}
\vspace{0.5cm}

We will focus on the case $f(R,\mathcal{G})=\alpha R^{n}\mathcal{G}^{m}$, which is related to the presence of the Noether Symmetries~\cite{capozziello2014} and represents a double inflationary scenario (see~\cite{DeLaurentis2015b} for more details).

From Eqs.~\eqref{var} and \eqref{mer}, we get:
\begin{eqnarray}
b&=&n-1\\
x_3&=&-nx_2\\
x_6&=&\frac{x_5}{(x_3-1)}\left\{\frac{nx_1}{b}+\frac{(m-1)}{(x_3-1)}\left[2(x_3-2)^2+\frac{x_1x_3}{b}\right]\right\}.
\end{eqnarray}
Hence, by using these relations and the constraint \eqref{constraint},
we can eliminate the equations for $x_1$, $x_2$ and $x_6$ from our
autonomous system, keeping a set of only four equations:
\begin{eqnarray}
\frac{dx_3}{dN}&=&\frac{x_1x_3}{n-1}-2x_3(x_3-2),\nonumber\\
\frac{dx_4}{dN}&=&-2x_3x_4-x_1x_4,\nonumber\\
\frac{dx_5}{dN}&=&\frac{x_5m}{x_3-1}\left[\frac{x_1x_3}{b}+2(x_3-2)^2\right]+\frac{x_1x_5}{n-1}-2x_5(x_3-2),\label{sys3}\\
\frac{dx_7}{dN}&=&-x_7(2x_3+x_1-1),\nonumber
\end{eqnarray}
where
\begin{eqnarray}
x_1&=&\frac{-1+x_3\displaystyle\frac{n+1}{n}+x_4+x_7+x_5(m-1)\left[\displaystyle\frac{1}{m-1}-2\displaystyle\frac{(x_3-2)^2}{(x_3-1)^2}\right]}{1+\displaystyle\frac{\left[n(x_3-1)+(m-1)x_3\right]x_5}{b(x_3-1)^2}}\;.
\end{eqnarray}

\subsubsection*{Fixed Points}

The fixed points can be obtained by setting the equations of the system \eqref{sys3} equal to zero, with $m=1-n$. The coordinates of the fixed points $(x_1,x_2,x_3,x_4,x_5,x_6)$ are:
\begin{eqnarray}
P_1: &~&\left(\frac{-4n+1}{n},\frac{-2n+2}{n^2},\frac{2n-2}{n},\frac{-5n^2+4n+2}{n^2},0,0\right), \quad \Omega_m=0, \quad \omega_{eff}=-\frac{3n-4}{3n},\nonumber \\
P_2: &~&\left(\frac{2n^2+2n-4}{2n^2-3n-1},\frac{-4n^2+5n}{2n^3-3n^2-n},\frac{4n^2-5n}{2n^2-3n-1},0,0,0\right), \Omega_m=0, \quad \omega_{eff}=-\frac{6n^2-7n+1}{6n^2-9n-3},\nonumber \\
P_3: &~&\left(0,0,0,1,0,0\right), \quad \Omega_m=0, \quad \omega_{eff}=\frac{1}{3},\nonumber \\
P_4: &~&\left(1,0,0,0,0,0\right),\quad \Omega_m=2, \quad \omega_{eff}=\frac{1}{3},\nonumber \\
P_5: &~&\left(-\frac{(3n-3)}{n},\frac{3-4n}{2n^2},\frac{4n-3}{2n},0,0,0\right), \quad \Omega_m=\frac{-8n^2+5n+3}{2n^2}, \quad \omega_{eff}=\frac{1-n}{n},\nonumber \\
P_6: &~&\left(0,-\frac{2}{n},2,0,\frac{-2-n}{n},0\right),\quad \Omega_m=0, \quad \omega_{eff}=-1,\nonumber \\
P_7: &~&\left(-4(2n-1)(n-1),0,0,0,\frac{8n^2-12+3}{8n^2-12n-1},\frac{4n(2n-3)(8n^2-12n+3)}{8n^2-12-1}\right),\nonumber \\
&~& \quad \Omega_m=0, \quad \omega_{eff}=\frac{1}{3},\nonumber \\
P_8: &~&\left(-1,0,0,0,0,0\right),\quad \Omega_m=0, \quad \omega_{eff}=\frac{1}{3}.\nonumber
\end{eqnarray}

Note that in most cases (with the exception of points $P_{3,4,8}$) the coordinates of the fixed points are dependent of the value of the parameter $n$. However,
there are values of $n$ for which the fixed points acquire an asymptotic character: $n=0,1,\frac{3\pm \sqrt{17}}{4}, \frac{3\pm \sqrt{11}}{4}$. 
The first one, namely $n=0$, corresponds to a Gauss-Bonnet gravity~\cite{Li}~\footnote{Note that, in this case, we should consider the action as $S=\int{d^4x\sqrt{-g}\left(\frac{R+f(G)}{2\kappa}\right)+\mathcal{L}_m}$.}.
The case $n=1$ correspond to Einstein term, for which this set of variables is useless. Finally, the cases where 
$n=\frac{3\pm \sqrt{17}}{4},\frac{3\pm \sqrt{11}}{4}$ are not allowed. Therefore, for our purposes these values of $n$ can be excluded.

We can see that $P_3$ is the standard radiation point and the point $P_1$ is a new radiation era which contains non-zero geometrical dark energy ($x_{1,2,3}\neq 0$). We should also note that the effective equation of state is well constrained by nucleosynthesis to be close to $1/3$ (at the radiation epoch)~\cite{Amendola2007} and, therefore, we can accept the point $P_1$ as a radiation era just for $n$ close to $1$.

Only the points $P_6$ and those originating from $P_{2,5}$ can be accelerated and only $P_4$ and $P_5$ can give rise to matter eras. The fixed point $P_4$ is characterised by $\omega_{eff}=1/3$ which can be ruled out as a correct matter era, since it presents the $a\propto t^{1/2}$ behaviour which means a ``wrong" matter era~\cite{Amendola2007}. On the other hand, the solutions which may give rise to a standard matter era, which means $\omega_{eff}=0$ ($a\propto t^{2/3}$), exist only for $n=\frac{1}{6}$ ($P_2$) or $n=1$ ($P_5$). Nevertheless, the case $P_2$ corresponds to $\Omega_m=0$, so this point does not give rise to a matter era dominated by a non-relativistic fluid, and the point $P_5$ seems to be like a matter era exclusively for $n$ close to $1$. 

In addition, the fixed point $P_2$ has an effective equation of state whose value depends upon $n$. We see that the condition for acceleration ($\omega_{eff}<-1/3$) is valid when $n<\frac{1}{4}(3-\sqrt{17})$ and $n>\frac{1}{4}(3+\sqrt{17})$. The behaviour of $\omega_{eff}$ as a function of $n$ is showed in Fig.\ref{weff_beh}. 

\begin{figure}[!htb]
    \centering
    \includegraphics[scale=0.5]{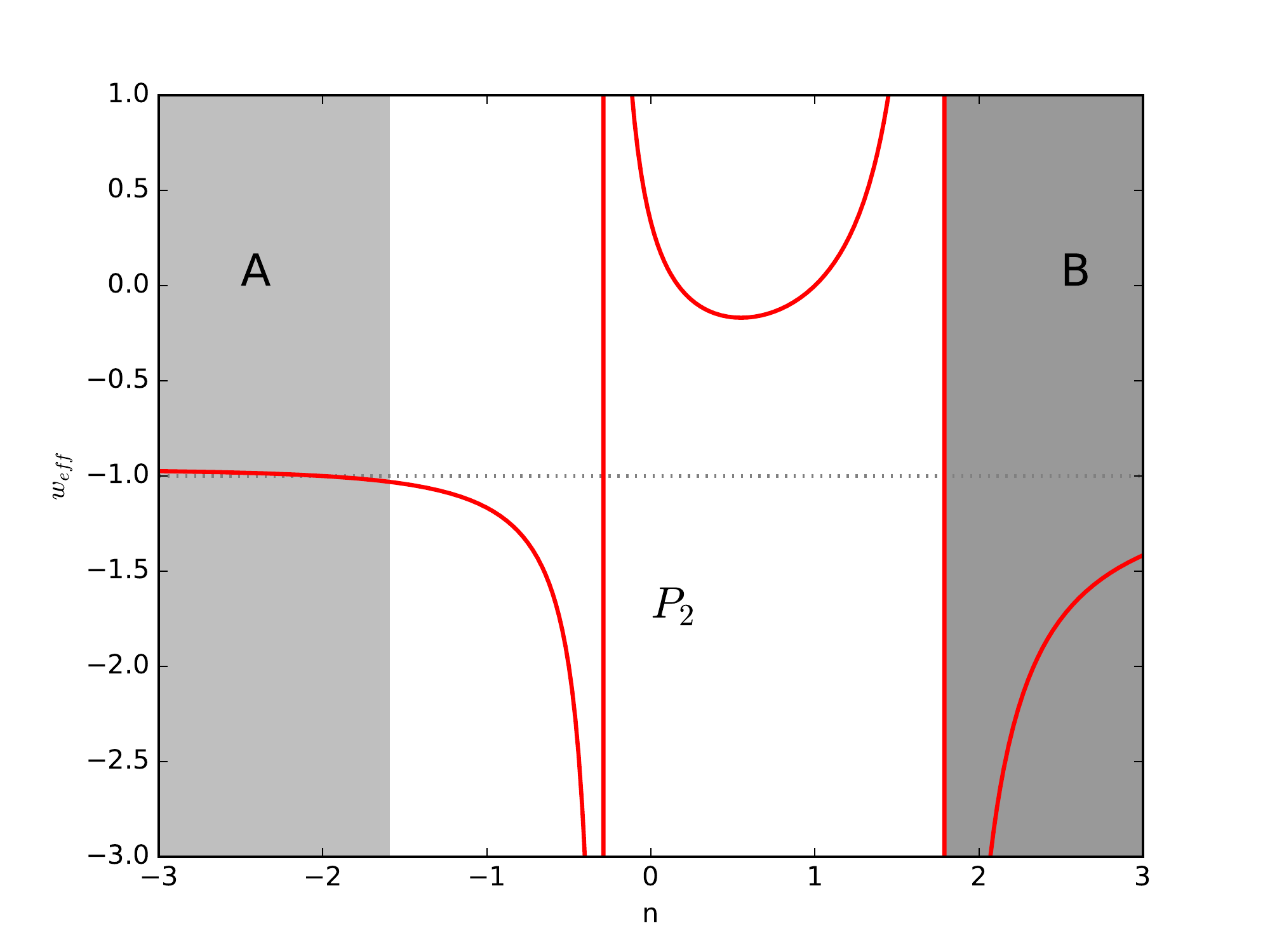}
    \caption{The effective equation of state $w_{eff}$ for $P_2$ as a function of $n$. This point is stable and accelerated in the grey regions. In the region (A) $n < -1.59$ the point is slightly phantom with $-1.03 < w_{eff} < -1$ and in the region (B) it is strongly phantom with $w_{eff} < -1.4$.}
    \label{weff_beh}
  \end{figure}
  
\subsubsection*{Stability}

The stability of the fixed points can be found using very well known techniques, 
which involve to linearize the dynamical equations around the equilibrium points and to find the eigenvalues of the
corresponding linearisation matrix (the Jacobian). Provided the Jacobian is well defined, we can classify
the fixed points according to the sign of the real part of the eigenvalues either as: stable nodes or attractors (all non zero eigenvalues with negative real part), unstable nodes or repellers (all non zero eigenvalues with positive real part), or saddle points (some non zero eigenvalues positive and some negative)\footnote{An eigenvalue with zero real part means that stability is indifferent along a given 1-dimensional sub-manifold of the phase space; we do not consider such eigenvalues in our analysis.}. The results are summarised in Tables \eqref{tab1} and \eqref{tab2}.

The fixed point $P_2$ behaves like a pure stable node for $n<-1.59$, $\frac{1}{4}(3-\sqrt{17})<n<\frac{1}{5}(2+\sqrt{14})$ and $n>\frac{1}{4}(3+\sqrt{17})$. For all other values of $n$ this point is a saddle. Hence, the point $P_2$ is both stable and accelerated just for $n<-1.59$ or $n>\frac{1}{4}(3+\sqrt{17})$. The standard radiation point $P_3$ behaves like a pure stable node for $n>\frac{1}{5}(2+\sqrt{14})$. For all other values of $n$ this point is a saddle. Although the eigenvalues of the new radiation point $P_1$ depends on the value of the parameter $n$, it always has a saddle-node character whatever the value of $n$.  The fixed points $P_4$ and $P_5$ are stable when $\frac{3}{4}<n<\frac{1}{5}(2+\sqrt{14})$ and have a saddle character otherwise. The fixed point $P_6$, or de-Sitter point ($\Omega_m=0$ and $\omega_{eff}=-1$), is stable when $n<\frac{1}{5}(2+\sqrt{14})$ and a saddle point otherwise.  The fixed point $P_7$ has always a saddle character, whatever the value of the parameter $n$. Finally, the fixed point $P_8$ can be either an unstable node, for $n>\frac{5}{4}$, or a saddle for all other values of the parameter $n$.

In Fig.~\ref{omega_evol}, we plot the evolution of the various quantities for $n=-2$ and $n=3.5$, respectively. For $n=-2$ (upper panels) we just have a ``bump" in matter density but if we look to the behavior of $w_{\mathrm{eff}}$ we can see a rapid transition from the radiation epoch to an accelerated epoch. Although, for $n=3.5$ (bottom panels), we have a peak in matter density, it is not possible to say which component is dominant. Instead, we must observe the behaviour of $w_{\mathrm{eff}}$ and note that a matter-dominated era is practically absent, since the model has a radiation-dominated epoch with a rapid transition to an accelerating universe with $w_{\mathrm{eff}}<-1$. 

In addition, we see in the stability analysis (cf. Table \eqref{tab2}) that in the range $\displaystyle {3}/{4}<n<(2+\sqrt{14})/5$, the matter-like points $P_4$ and $P_5$ behaves like stable nodes and the system cannot evolve to an accelerating epoch. This is better visualised in Figs. (3) and (4). The first one shows the behaviour of $\Omega_m$ and $w_{eff}$ as a function of $n$. The only possibility for $P_5$ to be a matter-like era is for $n$ close to $1$ and $w_{eff}$ close to $0$. However, at these values this point is an attractor (region A of Fig.(3)). Thus, the system cannot evolve to the accelerated stage. 
Figure (4) in turn shows such behaviour through the evolution of density parameters and the effective equation of state for $n=0.9$ and $n=0.98$, respectively. For $n=0.9$ (upper panels) the system evolves to the attractor $P_4$, with $\displaystyle \Omega_{\mathrm{m}}= 2$ and $\displaystyle w_{\mathrm{eff}}= {1}/{3}$. On other hand, when $n=0.98$ (bottom panels) the system evolves to the attractor $P_5$, characterised by an universe with $\sim10\%$ of matter and $\sim90\%$ of an accelerating component and with $w_{\mathrm{eff}}\approx 0.02$.

\begin{figure}
    \centering
    \includegraphics[scale=0.4]{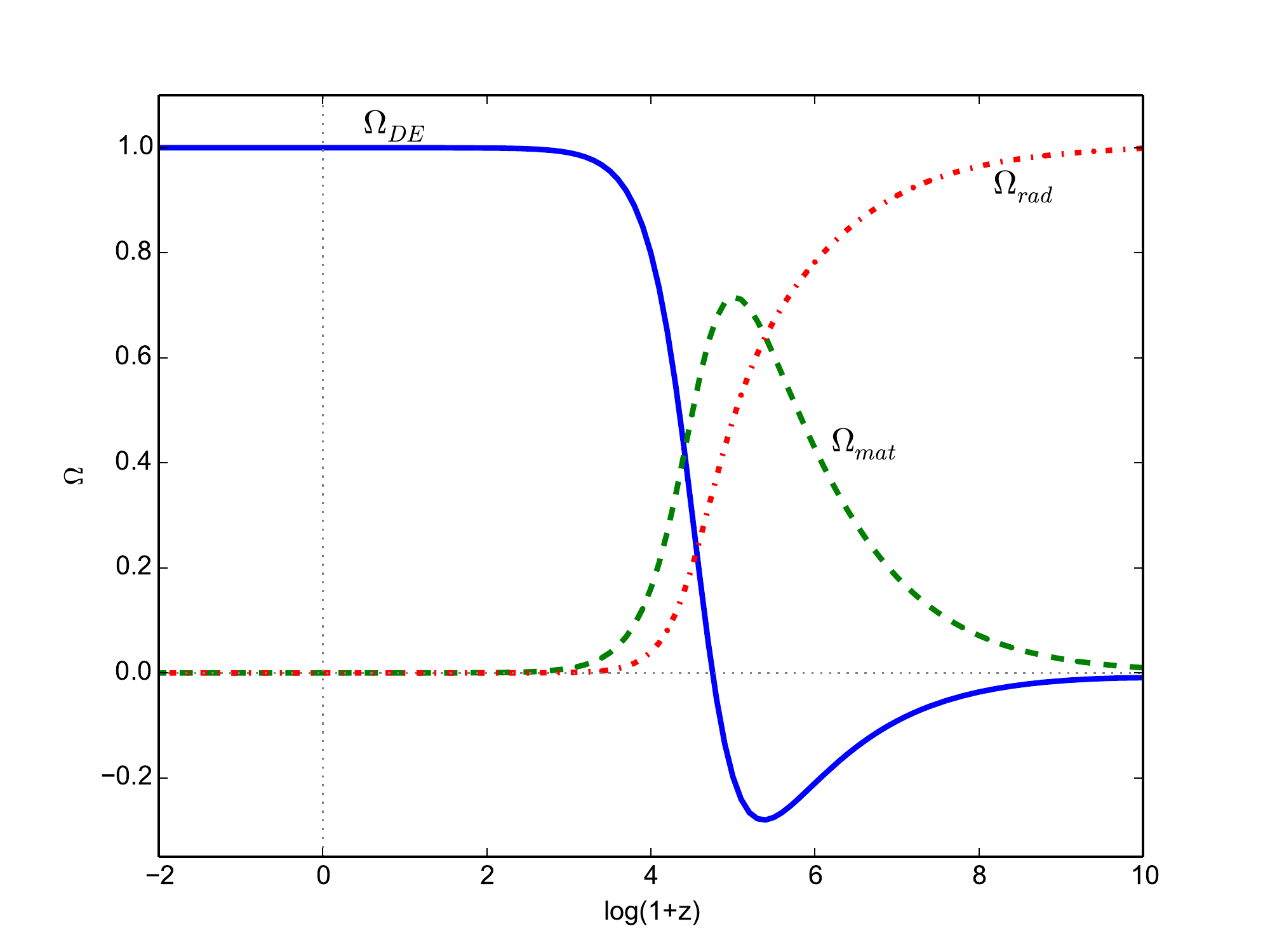}
    \includegraphics[scale=0.4]{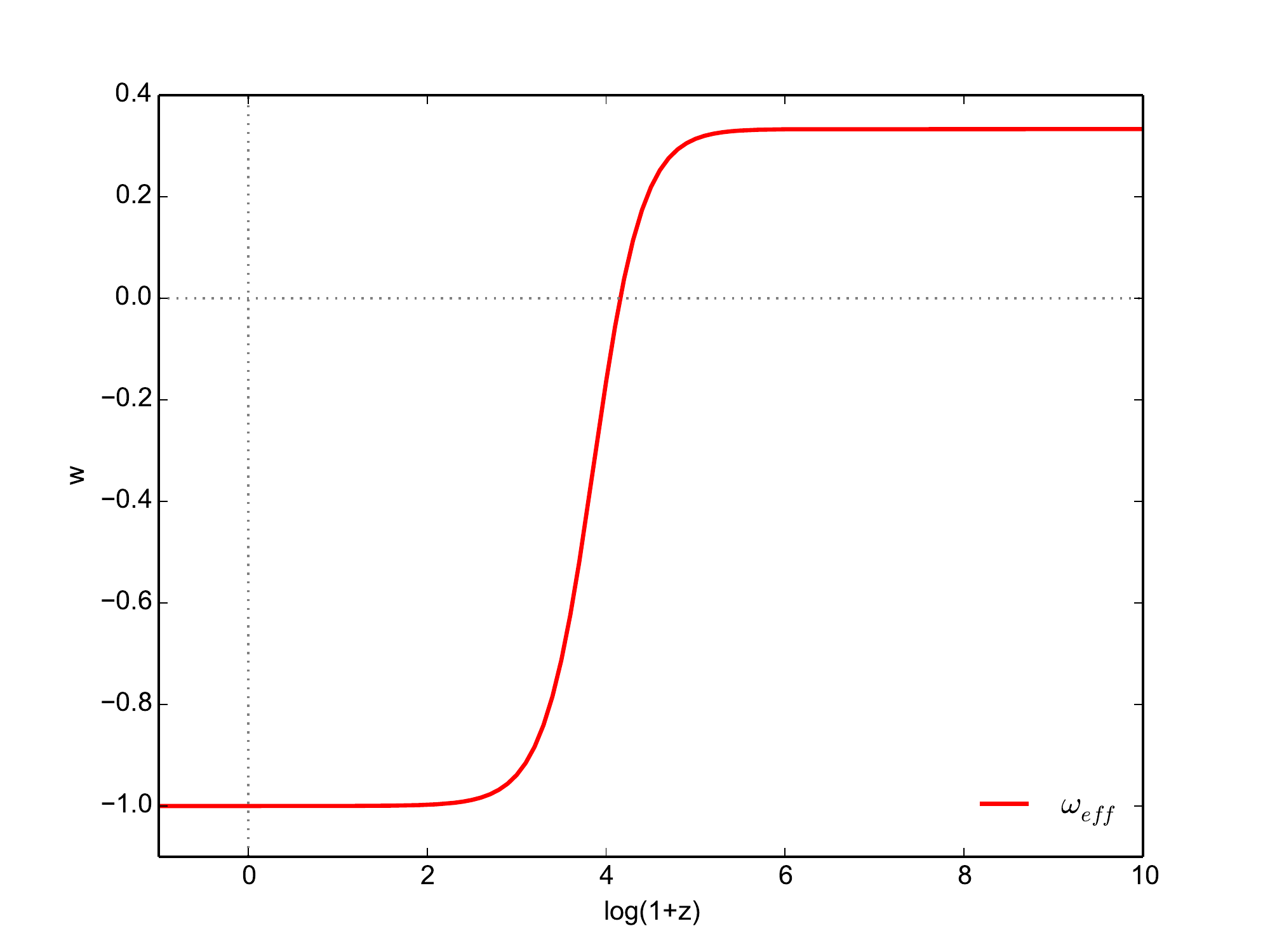}
    \includegraphics[scale=0.4]{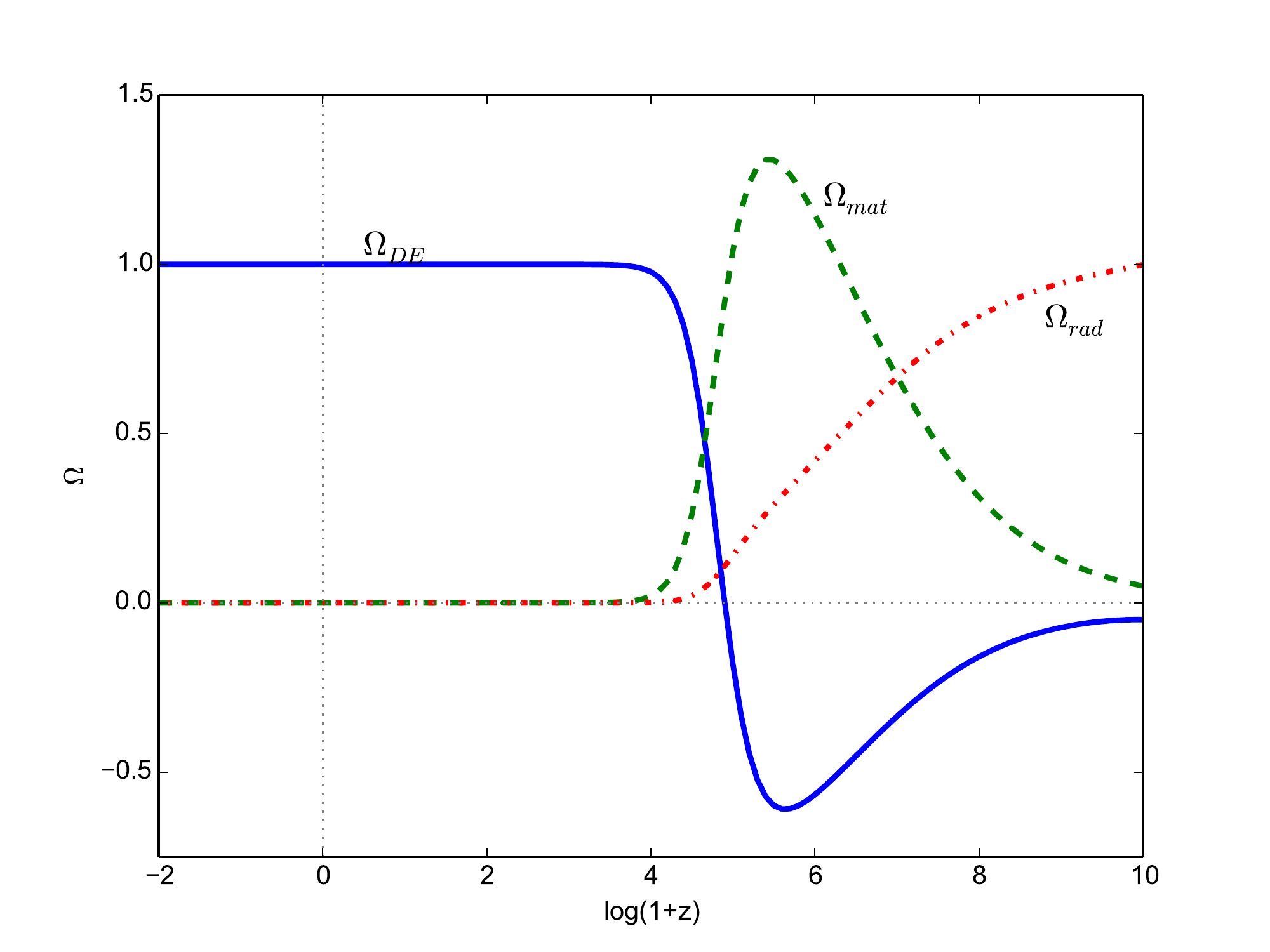}
    \includegraphics[scale=0.4]{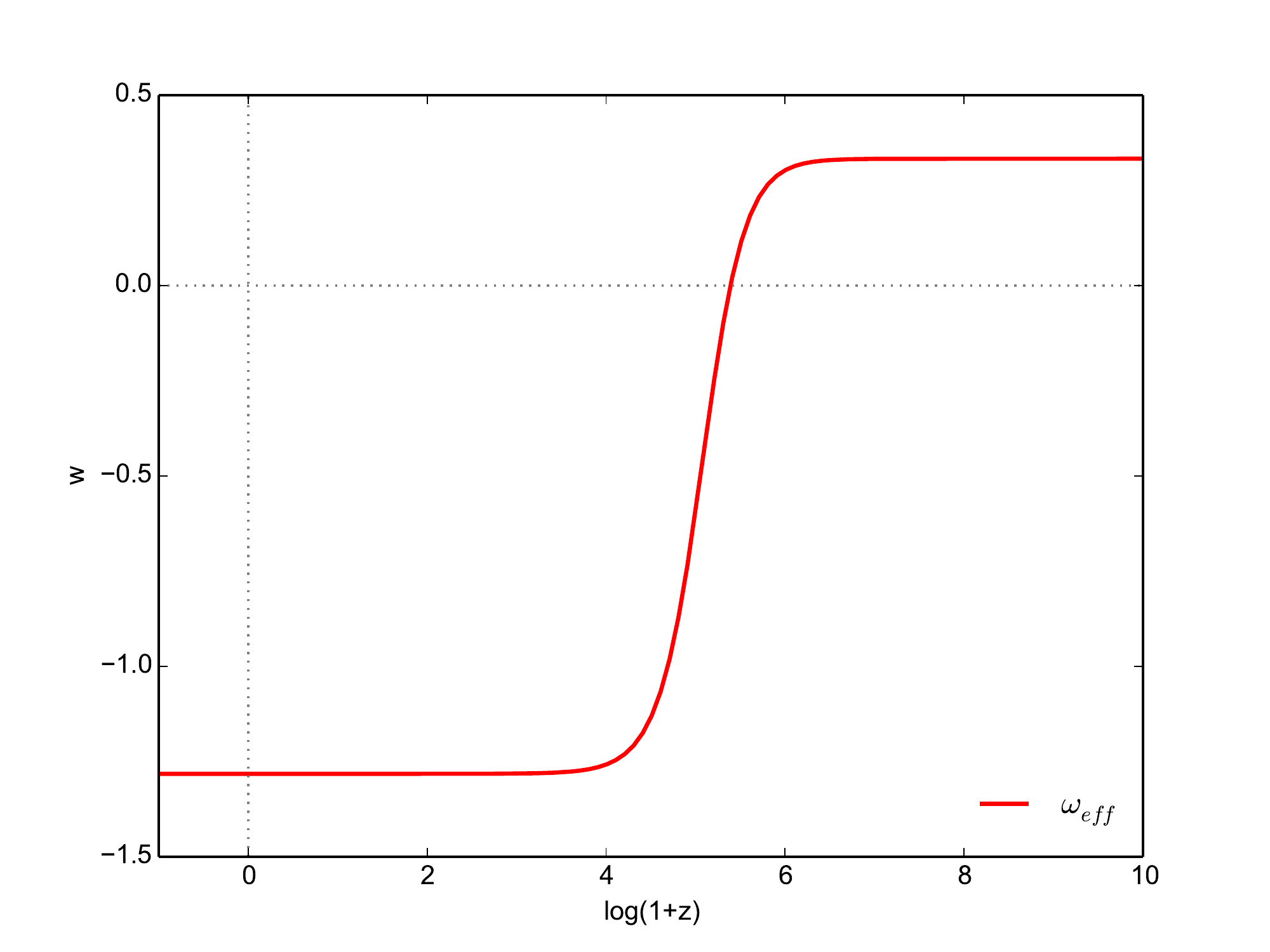}
    \caption{{\bf Upper panels:} The cosmic evolution of the density parameters for the model $f(R,G)=\alpha R^nG^{1-n}$, with $n=-2$. {\bf Bottom panels:} The same as in the upper panels but for $n=3.5$. Both of them consider the initial conditions $x_1=4.9\times10^{-3}$, $x_2=-10^{-8}$, $x_3=10^{-9}$, $x_4=0.999$, $x_5=x_6=0$ at the redshift $z=2.2\times 10^{4}$. }
    \label{omega_evol}
  \end{figure} 

\begin{figure}[!htb]
    \centering
    \includegraphics[scale=0.5]{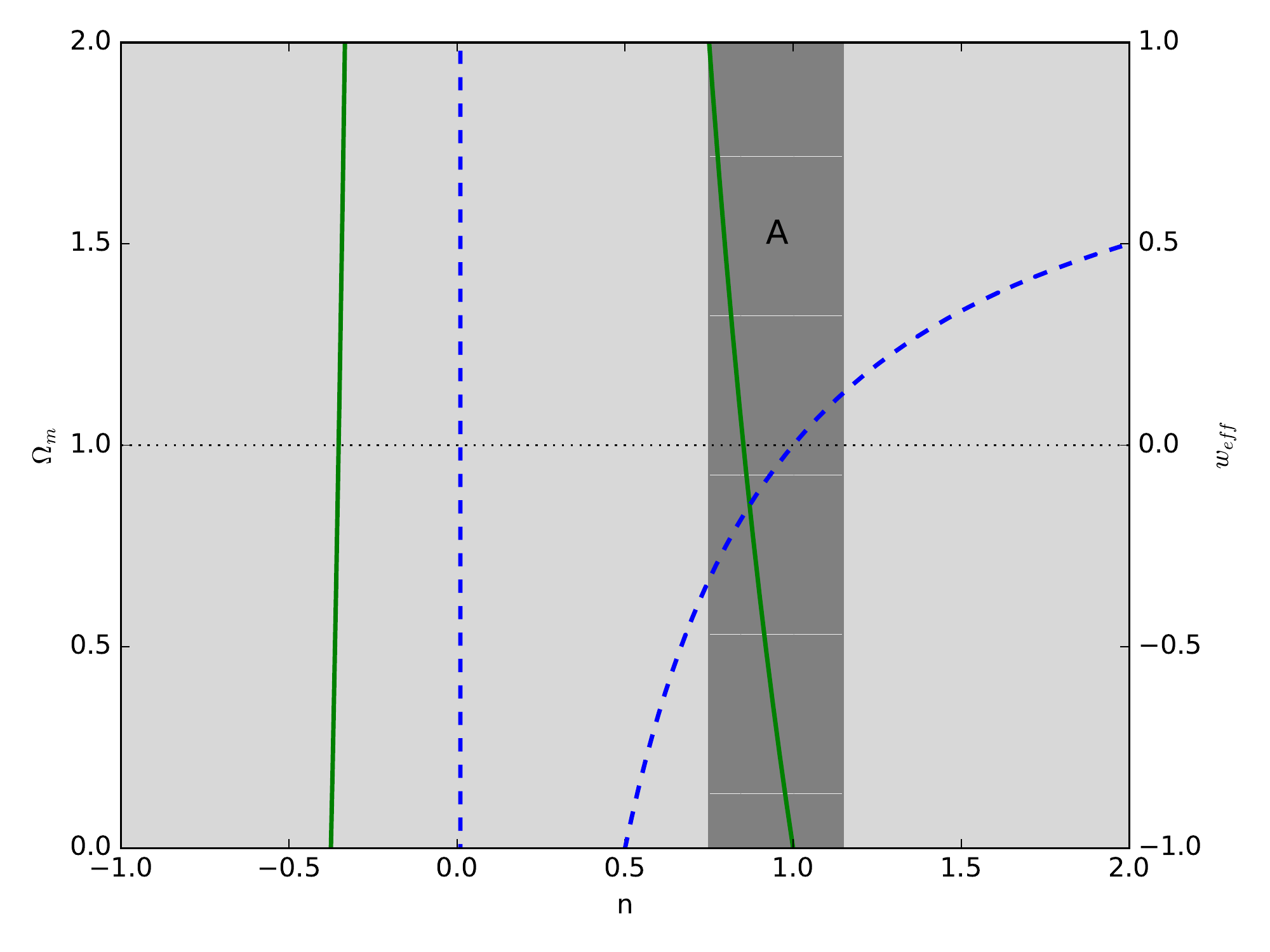}
    \caption{The matter density and the effective equation of state parameters, $\Omega_m$ (solid green line) and $w_{eff}$ (dashed blue line), respectively, for $P_5$ as a function of $n$. This point is an attractor (stable) when $\displaystyle {3}/{4}<n<(2+\sqrt{14})/5$ (dark grey region $A$) and a saddle otherwise (light grey regions).}
    \label{omega_weff}
  \end{figure}

\begin{table}
\begin{center}
\caption{\small{The eigenvalues associated with the fixed points in $f(R,\mathcal{G})=\alpha R^n\mathcal{G}^{1-n}$.}}
\vspace{0.2cm}
\begin{tabular}{cc}
\hline 
Point & Eigenvalues  \\
\hline   
$P_1$ & $[0,1,-\frac{4n-4}{n},\frac{8n^2-11n+4}{n(n-1)}]$ \\ 
      
$P_2$ & $\left[-\frac{10n^2-8n-4}{2n^2-3n-1},-\frac{8n^2-5n-3}{2n^2-3n-1},-\frac{2n(4n-5)}{2n^2-3n-1},-\frac{4n^3-7n+5}{2n^3-5n^2+2n+1}\right]$   \\
 
$P_3$ & $\left[0,1,4,\frac{8n^2-12n+5}{n-1}\right]$ \\
$P_4$ & $\left[0,-1,\frac{4n-3}{n-1},\frac{8n^2-12n+5}{n-1}\right]$ \\
$P_5$ &  $\left[0,-1,\frac{3-n}{2n^2-n-1},-\frac{4n-3)}{n},\frac{6n^2-8n+3}{n(n-1)}\right]$ \\
$P_6$ & $\left[-3,\frac{1}{n-1},-4,-4\right]$ \\
$P_7$ & $\left[-8n+8,8n^2-12n+4,8n^2-12n+5,\frac{8n^2-12n+5}{n-1}\right]$ \\
$P_8$ & $\left[1,2,\frac{4n-5}{n-1},\frac{8n^2-12n+5}{n-1}\right]$ \\
 \hline
\end{tabular}
\label{tab1}
\end{center}
\end{table}

\begin{table}
\begin{center}
\caption{\small{Stability of the fixed points for $f(R,\mathcal{G})=\alpha R^n\mathcal{G}^{1-n}$.}}
\vspace{0.2cm}
\begin{tabular}{ccccc}
\hline 
 & $n<-1.59$ & $-1.59<n<\frac{1}{4}(3-\sqrt{17})$ & $\frac{1}{4}(3-\sqrt{17})<n<\frac{3}{4}$ & $\frac{3}{4}<n<\frac{1}{5}(2+\sqrt{14})$  \\
\hline   
$P_1$ & Saddle & Saddle & Saddle & Saddle \\       
$P_2$ & Attractor & Saddle & Attractor & Attractor \\ 
$P_3$ & Saddle & Saddle & Saddle & Saddle \\
$P_4$ & Saddle & Saddle & Saddle & Attractor \\
$P_5$ & Saddle & Saddle & Saddle & Attractor  \\
$P_6$ & Attractor & Attractor & Attractor & Attractor \\
$P_7$ & Saddle & Saddle & Saddle & Saddle \\
$P_8$ & Saddle & Saddle & Saddle & Saddle \\
\hline
& $\frac{1}{5}(2+\sqrt{14})<n<\frac{5}{4}$ & $\frac{5}{4}<n<\frac{1}{4}(3+\sqrt{17})$ & $n>\frac{1}{4}(3+\sqrt{17})$ \\
\hline   
$P_1$ & Saddle &  Saddle & Saddle \\       
$P_2$ & Saddle & Saddle & Attractor \\ 
$P_3$ & Repellor & Repellor & Repellor \\
$P_4$ & Saddle & Saddle & Saddle \\
$P_5$ & Saddle & Saddle & Saddle \\
$P_6$ & Saddle & Saddle & Saddle \\
$P_7$ & Saddle & Saddle & Saddle \\
$P_8$ & Saddle & Repellor & Repellor\\  
\hline
\end{tabular}
\label{tab2}
\end{center}
\end{table}

\begin{figure}
    \centering
    \includegraphics[scale=0.4]{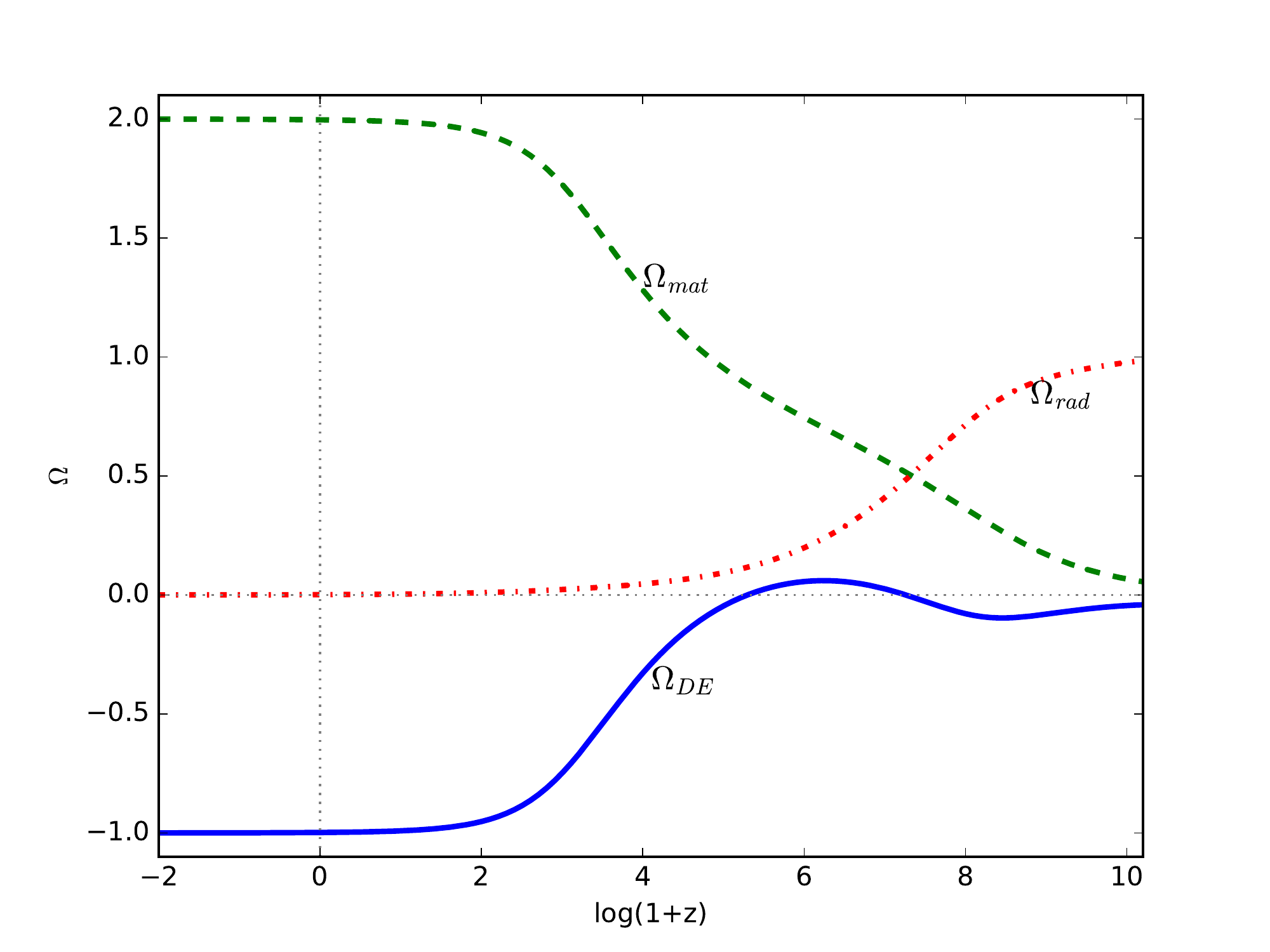}
    \includegraphics[scale=0.4]{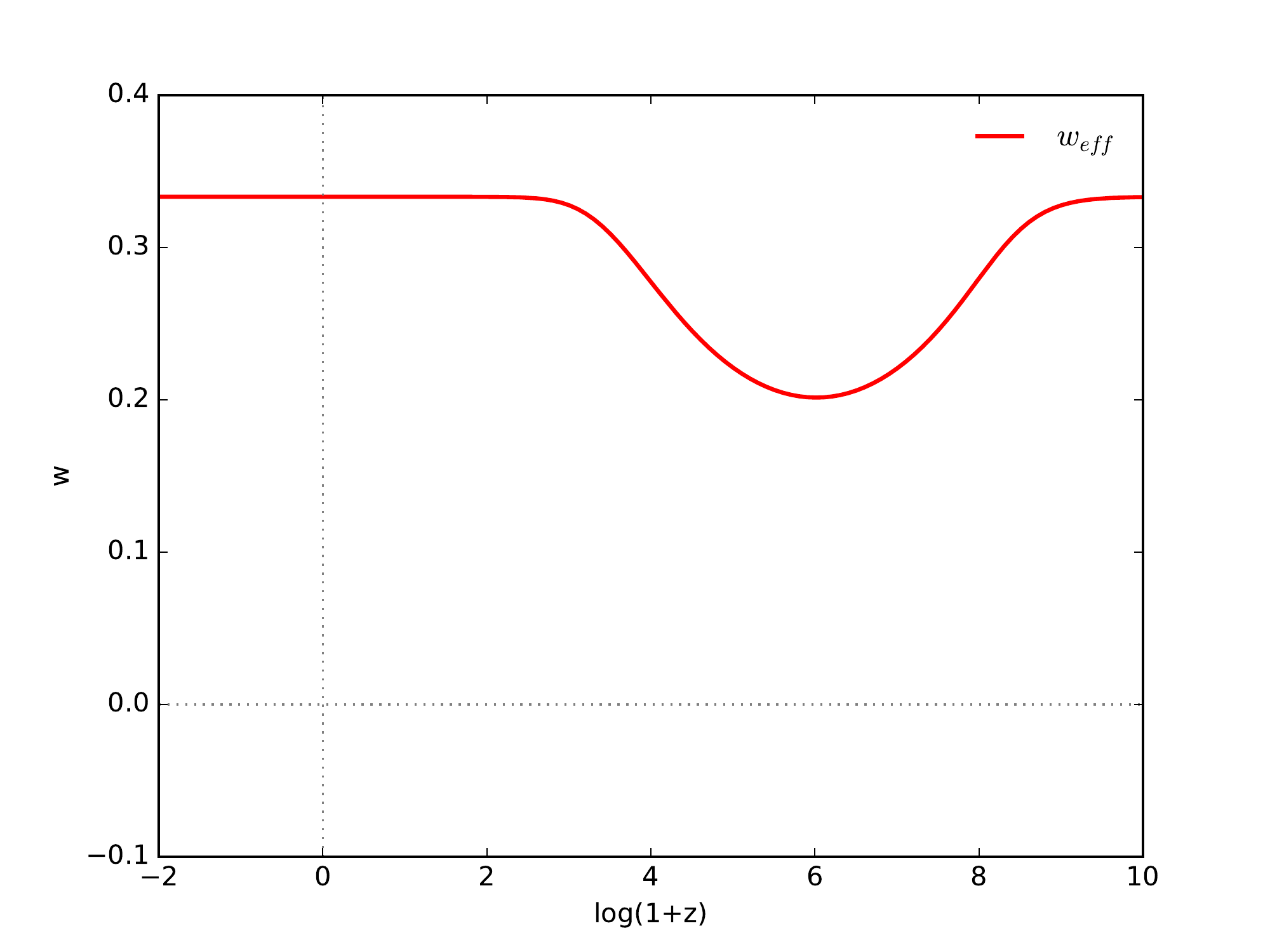}
    \includegraphics[scale=0.4]{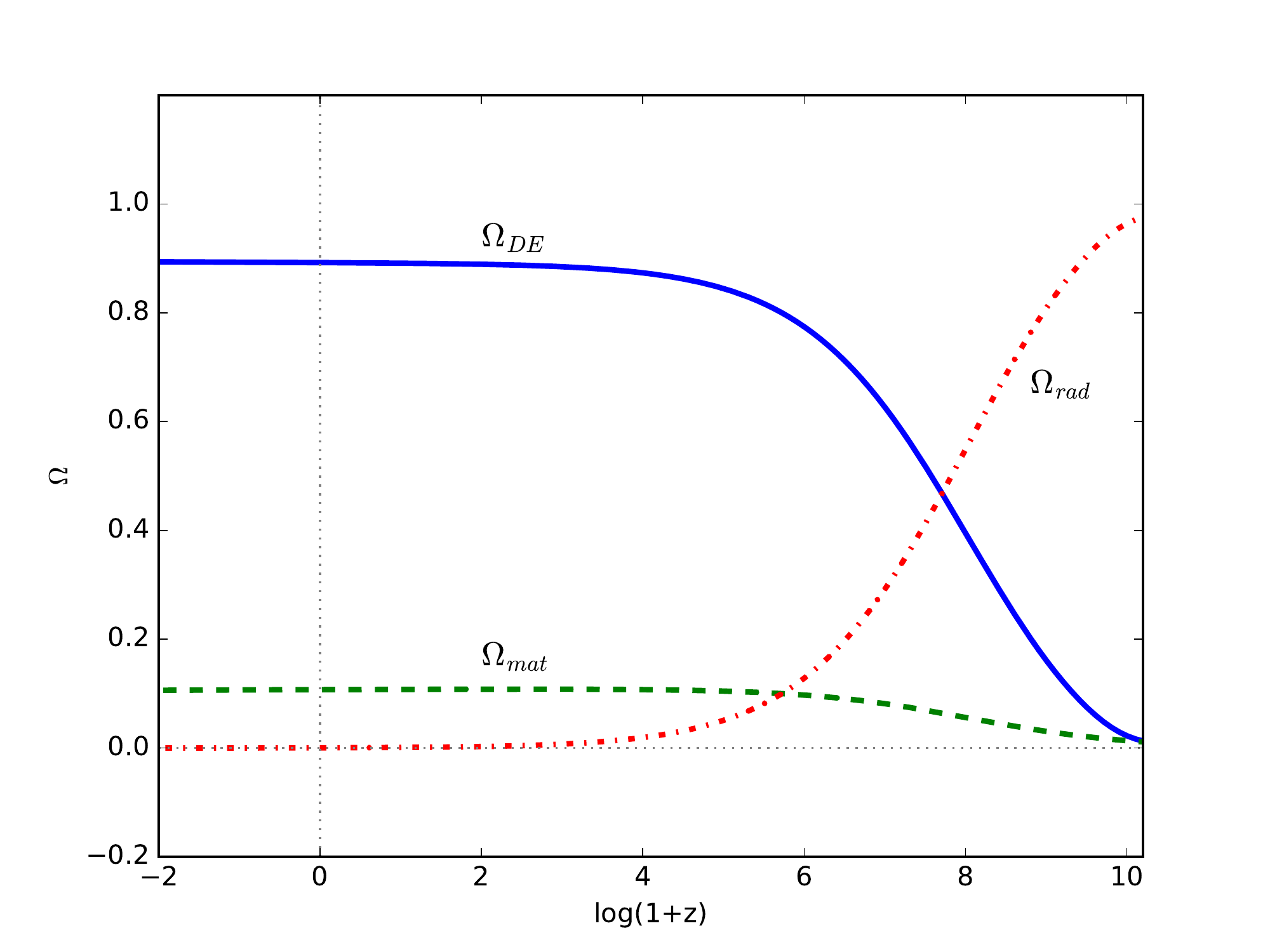}
    \includegraphics[scale=0.4]{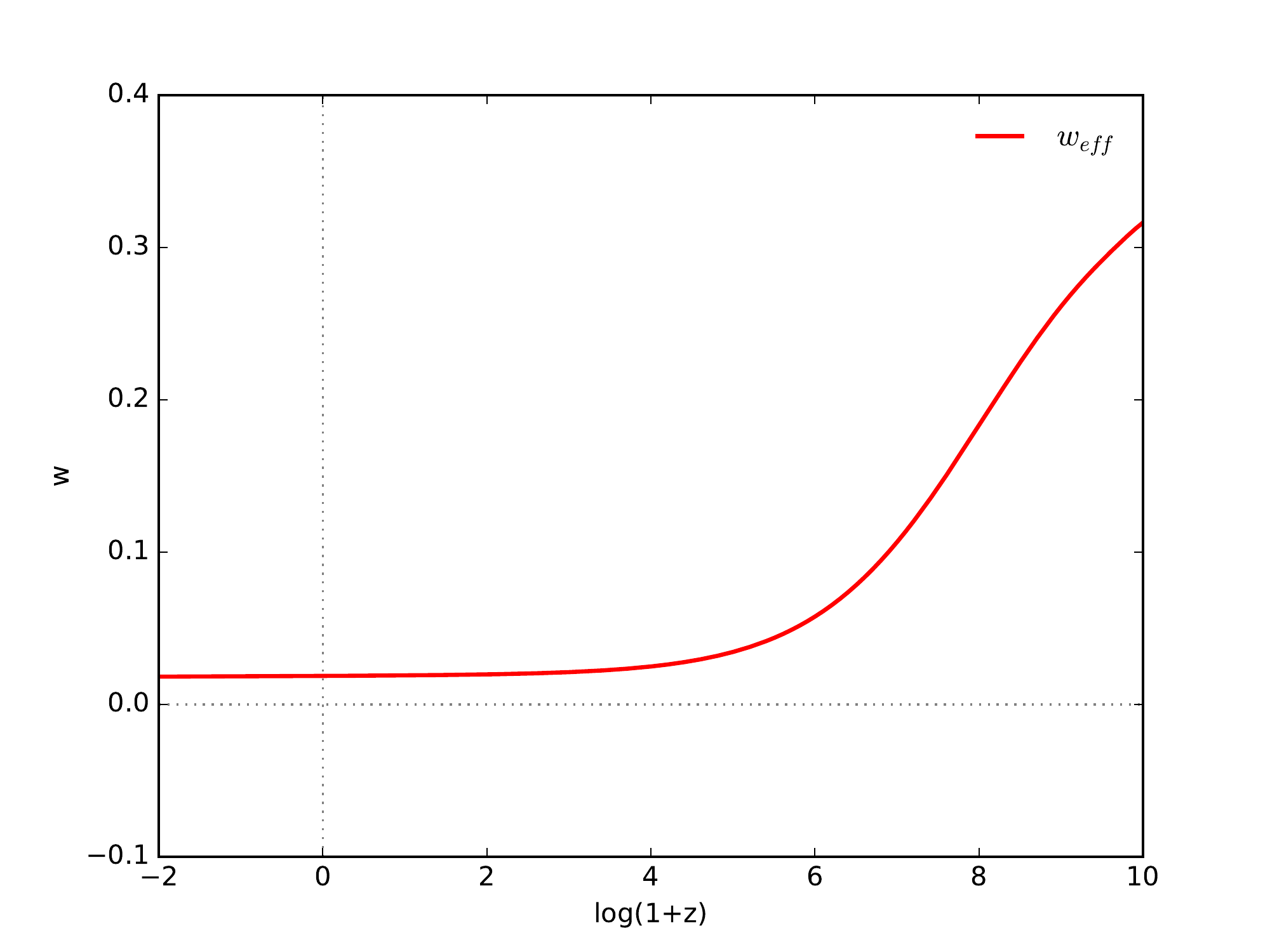}
    \caption{{\bf Upper panels:} The cosmic evolution of the density parameters of the model $f(R,G)=\alpha R^nG^{1-n}$ for $n = 0.9$, with the initial conditions $x_1 = 4.9 \times 10^{-2}$, $x_2 = -10^{-6}$, $x_3 = 10^{-4}$, $x_4 = 0.999$, $x_5 = x_6 = 0$ at the redshift $z = 3 \times 10^4$. 
    {\bf Bottom panels:} Same as in the upper panels, but for $n = 0.98$ and with the initial conditions $x_1 = 0$, $x_2 = 0$, $x_3 = 10^{-3}$, $x_4 = 0.98$, $x_5 = 0$, $x_6 = 0$ at the redshift $z = 3 \times 10^4$. In this case, the universe has a radiation dominated epoch and evolves to an attractor
corresponding to point $P_5$ with $10\%$ of matter density and $90\%$ of "dark energy'', with $\omega_{eff} = 0.02$. }
    \label{omega_evolt}
\end{figure}

\section{Discussion and conclusions}\label{sec5}

In this work we considered a generic case of $f(R,\mathcal{G})$ theories of gravity and applied a dynamical system analysis to investigate their cosmological features. As expected, when the variables $x_5=0$ and $x_6=0$, the phase space presents the same fixed points previously obtained for $f(R)$ gravity in Ref.\cite{Amendola2007}. However, the resulting general system is very difficult to analyze without specifying the function $\Gamma$ (i.e. the form of $f(R,\mathcal{G})$). Thus, we have focused our work on a specific type of function $f(R,\mathcal{G})$, which makes our analysis less general than ideal, but allowed us to analyze the stability and information about the global behavior of this cosmology.

We applied the DSA to a power-law class of fourth order gravity
model, namely $f(R,\mathcal{G})=\alpha R^n\mathcal{G}^{1-n}$, and we found some very interesting preliminary results for the finite phase space. There were found eight fixed points, listed in Table \eqref{tab1}. Two of them, specifically the points $P_2$ and $P_6$, have solutions that admit accelerated expansion, and so, can be considered as possible candidates able to model the dark energy era. The fixed point $P_6$ has an equation of state parameter equal to $-1$ and $\Omega_m=0$, so it can represent a de Sitter solution. From the stability point of view this point has a pure stable node character for $n<\frac{1}{5}(2+\sqrt{14})$ and, consequently, can be considered as a final attractor. On the other hand, for $n<-1.59$, $\frac{1}{4}(3-\sqrt{17})<n<\frac{1}{5}(2+\sqrt{14})$ and $n>\frac{1}{4}(3+\sqrt{17})$ the point $P_2$ represents an attractor and for all the other values of $n$ this point has a saddle character.

The point $P_3$ is the standard radiation phase and has unstable regions for $n>\frac{1}{5}(2+\sqrt{14})$. The point $P_1$, a new radiation era, has always a saddle character.

It is worth mentioning that the unique point which might give rise to the matter era is the point $P_5$, but exclusively for $n$ close to $1$ to produce $\omega_{eff}$ close to $0$. However, for this value of parameter $n$ this point behaves like a stable node and, hence, can not represent this epoch. In contrast, although the point $P_4$ presents $\Omega_m$ different from $0$, the effective equation of state is given by $\omega_{eff}=1/3$, which produces a ``wrong" matter era that can be ruled out, e.g. by the angular diameter distance of the CMB acoustic peaks~\cite{AmendolaDACMB}. Therefore, this class of models does not contain a standard matter era. 

Finally, the points $P_{7,8}$ cannot represent both matter eras ($\Omega_m=0$) or an accelerated phase ($\omega_{eff}=1/3$).

Therefore, the general features for this class of $f(R,\mathcal{G})$ model can be outlined as: 
\begin{itemize}
\item The $f(R,\mathcal{G})$ model has a standard radiation dominated epoch whatever the value of parameter $n$ ($P_3$).
\item The standard matter dominated epoch is present exclusively for $n\rightarrow 1$, which produces $\omega_{eff}$ close to $0$ ($P_5$). However, from the stability point of view, for this value of $n$ this point behaves like an attractor, and consequently, cannot represent the matter dominated epoch. 
\item A possible way to get a saddle matter point is for $n<\frac{3}{4}$ or $n>\frac{1}{5}(2+\sqrt{14})$ in $P_5$, but in this case $\omega_{eff}\neq 0$. So, it cannot represent a matter dominated epoch either.
\item The de-Sitter acceleration point $P_6$ exist for $n<\frac{1}{5}(2+\sqrt{14})$ and the new acceleration point $P_2$ exist for $n<-1.59$ and $n>\frac{1}{4}(3+\sqrt{17})$.
\end{itemize}

Concern the non-existence of a matter domination epoch, the authors of \cite{DeLaurentis2015b} have already considered this class of models to study the grow factor and noticed that for values of $n>1$, the universe accelerates forever without the possibility of structure formation. Finally, we have found that this particular class of models can not describe a reasonable cosmology given by any trajectory passing near by $P_3$ or $P_1$ (for $n\approx 1$), then passing by a matter dominated epoch and landing on an accelerated attractor. Thus, we emphasise  that it is necessary a more detailed study considering also observational data as well as more realistic models in the context of $f(R,\mathcal{G})$ gravity.

\section*{Acknowledgements}

S. Santos da Costa acknowledges financial support from Coordena\c{c}\~{a}o de Aperfei\c{c}oamento de Pessoal de N\'ivel Superior (CAPES).
F.V. Roig thanks the financial support from CAPES and Conselho Nacional de Desenvolvimento Cientifico e Tecnologico (CNPq). M. Benetti is supported by the Funda\c{c}\~{a}o Carlos Chagas Filho de Amparo \`{a} Pesquisa do Estado do Rio de Janeiro (FAPERJ - fellowship {\textit{Nota 10}}). S. Capozziello is supported in part by the INFN sezione di Napoli, iniziative specifiche TEONGRAV and QGSKY. J. Alcaniz acknowledges support from CNPq (Grants no. 310790/2014-0 and 400471/2014-0) and FAPERJ (grant no. 204282).  M. De Laurentis  is supported by ERC Synergy Grant Black Hole Imaging the Event Horizon of Black Holes awarded by the ERC in 2013 (Grant No. 610058). This article is based upon work from COST Action CA15117  "Cosmology and Astrophysics Network for Theoretical Advances and Training Actions" (CANTATA), supported by COST (European Cooperation in Science and Technology).

\end{document}